\newcommand{\half}{\frac{1}{2}}

\newcommand{\ds}{\displaystyle}
 \documentstyle[preprint,aps,tighten]{revtex}
\begin{document}
\draft
\title{Detailed Balance and Intermediate Statistics}
\author{R. Acharya $^a$ and P. Narayana Swamy $^b$}
\address{$^a$ Professor Emeritus Physics,
Arizona State University, Tempe AZ 85287\\
$^b$ Professor Emeritus Physics, Southern Illinois University,
Edwardsville IL 62026} \maketitle
\begin{abstract}

We present a theory of particles,  obeying intermediate statistics
(``anyons"), interpolating between Bosons and Fermions,
 based on the principle of Detailed Balance.
 It is demonstrated
 that the scattering probabilities of identical particles
 can be expressed in terms of the basic numbers, which arise
naturally and logically in this theory. A transcendental equation
determining the distribution function of anyons is obtained in
terms of the statistics  parameter, whose limiting values 0 and 1
correspond to Bosons and Fermions respectively.  The distribution
function is determined as a power series involving the Boltzmann
factor and the statistics
 parameter and we also express the distribution
function as an infinite continued fraction. The last form enables
one to develop approximate forms for the distribution function,
with the first approximant agreeing with our earlier
investigation.

\end{abstract}

 \vspace{2in}
Electronic address: $\, ^a$ Raghunath.acharya@asu.edu,
 $^b$ pswamy@siue.edu\\

\indent August 2003

 \indent 

\section{Introduction}

We  formulate a theory of particles  obeying intermediate
statistics, interpolating between  Bosons and Fermions, which
might be called anyons. Our formulation will be based on
 two assumptions: 1) The exchange symmetry or permutation
of the coordinates of the particles in the many particle wave
function is accompanied by multiplication by a complex number $f$,
thus generalizing the symmetric or antisymmetric wave functions
and 2) The principle of Detailed Balance: if $n_1, n_2$ represent
the average occupation numbers of states labelled by 1 and 2, then
the number of transitions flowing from 1 to 2 must equal that
flowing from 2 to 1 at equilibrium.

\vspace{.15in}

The particles described by this theory may or may not be the same
as the anyons obeying intermediate or fractional statistics
discussed in the literature. The objects named anyons
\cite{Leinnas,Wilczek,Forte,Chern,Lerda,Khare,Frau,Frappat} carry
both an electric charge and a magnetic flux. They have attracted a
great deal of attention and  have been the subject of intense
investigation in the literature. The anyons arise from the special
circumstance in 2+1 space-time dimensions, where the permutation
group is the braid group and the Chern-Simons theory provides a
natural realization of the anyons \cite{Wil,Jackiw}. There has
been a great deal of discussion in the literature on the
thermostatistics of anyons \cite{RAPN}.  Since the real world is
strictly in a 3+1 dimensional space, anyons may not be real
particles: they could be quasi-particles playing important roles
in condensed matter phenomena. More recently, the subject of
generalized statistics has been investigated in one dimension in
the context of many kinds of statistics \cite{Poly}. In contrast
to the theory of anyons familiar in the literature, our present
approach is not limited to two space dimensions and is valid in
the real world of 3+1 dimensional space-time. It is important to
point out that this is due to the fact that we do not invoke the
spin-statistics theorem and do not require ordinary spin to be
interpreted in two dimensional space.

\vspace{.15in}

On one hand, the theory based only on Detailed Balance does not
have the features of the full-fledged theory of anyons which takes
advantage of the braid group in two dimensions. On the other hand,
in this formulation we investigate the idea of interpolating
statistics only in the context of the statistical mechanics of a
gas in equilibrium, without the constraints imposed by
 quantum field theory. Here we shall use the name
anyons, for convenience, to refer to particles obeying
intermediate statistics, with the understanding that the present
work needs to be developed further by incorporating additional
assumptions before the connection with  true anyons described in
the literature could be established.

\vspace{.15in}

The subject of anyons has been well investigated in the
literature, especially in the context of quantum field theory and
the braid group \cite{Wilczek,Lerda,Frau,Frappat}. Interesting
results have also been derived to describe the thermostatistics of
anyons, such as determining the virial coefficients
\cite{RAPN,other}. However, the theoretical basis of the
statistical mechanics of anyons has not really been established.
For instance, the distribution function for the anyons has not
been determined in an exact form. We had used an ansatz for the
distribution function in our earlier work in order to derive many
of the thermodynamic properties \cite{RAPN}. More recently,
Chaturvedi and Srinivasan \cite{Chatur} have done a comparative
study of the different interpolations between Bose-Einstein (BE)
and Fermi-Dirac (FD) statistics available in the literature,  in
the context of the most general interpolations, including Haldane
statistics and Gentile statistics, with the conclusion that the
distribution function introduced by us \cite{RAPN} has some
desirable features.

\vspace{.15in}

Consequently it is worthwhile to investigate the basic theoretical
structure of anyons from the point of view of statistical
mechanics and investigate the distribution function of the anyons
if possible. This is the goal of this paper. Not only shall we
determine the distribution function of the particles obeying the
interpolation statistics, we shall also  formulate a theory which
leads to this determination without resorting to the restriction
to a two dimensional space. Our formulation will be based on the
quantum theory of many particles permitting a generalized
interpolating exchange symmetry statistics with no other
assumptions. Remarkably, this formulation leads to an exact theory
which requires the employment of basic numbers.

\vspace{.15in}

In Sec.II we study the scattering probability of many particle
states obeying interpolating statistics. We establish the fact
that the basic numbers arise naturally and logically in this
theory. We introduce the method of Detailed Balance in Sec. III
and derive a transcendental equation for the distribution function
in terms of the Boltzmann factor and the statistics determining
parameter. The distribution function of anyons is studied in Sec.
IV where we show that a closed form solution is not possible. We
present the exact solution for the distribution function as an
infinite series as well as in the form of an infinite continued
fraction which is amenable to approximations. Sec. V contains a
brief summary and conclusions.

\vspace{.15in}

 Let us introduce an ensemble of particles, the
anyons,  which obey a generalized statistics, interpolating
between  (BE) and
 (FD) statistics. We begin with the framework for building the wave functions
 of these anyons by a
 generalized procedure of  $f$-symmetrizing in such a way that
  it will reduce, in appropriate limits,
   to the standard procedure of symmetrizing for
  Bosons
  and anti-symmetrizing for Fermions. The operation of
permutation or exchange of the coordinates of the many particle
wave function results in multiplication by the complex number $f$,
the exchange symmetry parameter, so that

\begin{equation}\label{1}
{\cal P} \Psi_n(\cdots , q_i, \cdots , q_j, \cdots) = f \Psi_n
(\cdots , q_j, \cdots , q_i, \cdots)\, .
\end{equation}
Since the Hamiltonian has the property ${\cal P}^{-1}{\cal H}
{\cal P} = {\cal H}$, it follows that ${\cal H} ({\cal P} \Psi_n)
= E_n ({\cal P} \Psi_n)$ and thus ${\cal P} \Psi_n$ is an
eigenfunction of the Hamiltonian with the same eigenvalue as
$\Psi_n$. Thus $f \Psi_n$ is proportional to $ \Psi_n$ and
consequently we may take $|f|^2 \Psi_n= \Psi_n$.  Hence the
general exchange symmetry that would lead to an intermediate
statistics could be implemented by the complex number $f$ with the
property $|f|^2=1$. We shall accordingly choose the exchange
symmetry to be implemented by  $f = e^{i \pi \alpha}$, where
$\alpha$ is the statistics determining parameter, $0 \leq \alpha
\leq 1$, so that $f^{*}= f^{-1}$. The limits $f \rightarrow 1, -1$
correspond respectively to the BE and FD statistics representing
Bosons and Fermions. This procedure for incorporating exchange
symmetry among anyons is justified \cite{Frau} by the special
property of rotations in two dimensions. Here we may treat it as
an assumption or an ansatz, not restricted to two space dimensions
but valid in any number of dimensions.

\section{Many particles and Quantum probabilities}

Following Feynman \cite{Feyn}, we consider the two particle
scattering amplitude defined by the product $a_1\, b_2$, where $
a_1 = \langle 1 | a \rangle $ describes the scattering of particle
$a$ into state 1 and $ b_2 = \langle 2 | b \rangle $ describes the
process $b \rightarrow 2$. We shall take 1 and 2 to be the same
state at the end in order to deal with identical particles. The
exchange symmetry has to do with the process corresponding to $a
\rightarrow 2, \, b \rightarrow 1 $ which is indistinguishable
from the direct process and the amplitude for this process would
be $f \, a_2 \, b_1 $ due to the exchange factor $f$. The total
probability amplitude is the sum of the direct and exchange
processes.  Employing the abbreviation $\langle 1 | a \rangle \, =
\langle 2 | a \rangle = a $, we find the probability of this two
particle scattering process involving non-identical particles to
be
\begin{equation}\label{x1}
    p_{non}^{(2)}= (1 + |f|^2) |a|^2 |b|^2 \; = \; 2 (|a|^2 |b|^2)\, ,
\end{equation}
since $|f|^2 =1$. This probability is the same as for ordinary
Bosons \cite{Feyn}. However, for identical particles, we need to
take account of interference between the two processes and that
makes a great deal of difference. We obtain the probability in
this case to be
\begin{equation}\label{x2}
    p_{identical}^{(2)}= (1 + |f|^2 + f + f^{-1})\, |a|^2 |b|^2\, ,
\end{equation}
since $f^{*}= f^{-1}$ and this  probability depends on the
statistics determining parameter $\alpha$.  In the limit $f
\rightarrow 1$, this would reduce to the case of Bosons and it
would be twice as much as in Eq.(\ref{x1}) for the non-identical
particles. For arbitrary $f$  the probability for the process
involving identical particles relative to that for non-identical
particles is given by
\begin{equation}\label{x3}
    P_{identical}^{(2)}=  \half \left
    ( \, 2+f+f^{-1}\,\right )\,|a|^2 |b|^2 .
\end{equation}
In what follows, we shall omit $|a|^2, |b|^2$ etc. since the
single particle states are properly normalized. Understanding this
to be a proportionality, we might henceforward refer to this
itself as the probability for the process, or just the probability
of the two particle state,  for simplicity.

\vspace{.15in}

   Similarly, we consider the  three-particle processes
   $a \rightarrow 1, b \rightarrow 2, c \rightarrow 3$ together
   with the exchange processes with factor $f$ for each exchange
   operation, thus resulting in the combination
   $abc + f\, acb + f\, bac + f^2 \, bca + f^2 \, cab + f^3 \,
   cba$ with the same abbreviation as earlier. We can now
   determine the probability for
   the three particle process, or just for the three particle state
   with
   identical particles, as
\begin{equation}\label{5}
    P^{(3)}= \frac{1}{6} (1 + 2f^{-1} + 2 f^{-2} + f^{-3})
    (1 + 2f + 2 f^2 + f^3 )\, .
\end{equation}
We can reduce this to the form
\begin{eqnarray}
  P^{(3)} &=& \frac{1}{6} \left \{10 + 8 (f + f^{-1})
  + 4 (f^2 + f^{-2} ) + ( f^3 + f^{-3})  \right \}\nonumber\\
   &=& \frac{1}{6} \left  \{
  6 + 7 (f+ f^{-1}) + 4 (f^2 +f^{-2} +1) +
  (f^3 + f^{-3}+ f + f^{-1})   \right \}\, .\label{5A}
\end{eqnarray}
Recognizing the pattern here, we observe that the right hand side
in the expressions for the probabilities in Eqs.(\ref{x3}),
(\ref{5}), (\ref{5A}) contain basic numbers \cite{Exton} , with
the base $f$. They are indeed expressed succinctly in terms of the
basic numbers defined by
\begin{equation}\label{6}
    [n]_f= \frac{f^n - f^{-n}}{f-f^{-1}}\, .
\end{equation}
 We shall henceforward omit the subscript
$f$ for simplicity. Here $f=e^{i \pi \alpha }$ , the BE limit is
$\alpha \rightarrow 0, f \rightarrow 1 $ and the FD limit is $
\alpha \rightarrow 1, f \rightarrow -1 $ and $f^* = f^{-1}$. Our
formulation is symmetric under $f \rightarrow f^{-1}$ and is the
familiar generalization of the basic numbers introduced long ago
by F. H. Jackson \cite{Exton}. Studying the limits we find that
the Bose limit gives $[n] \rightarrow n$ while the Fermi limit is
quite different: $[n]\rightarrow (-1)^{n+1}\, n$ which becomes
$-n$ for even numbers and $+n$ for odd numbers. In this limit, we
are therefore dealing with a generalization of ordinary fermions
for which the exclusion principle is not automatically valid: in
the FD limit it can be imposed by hand only. Returning to the
basic numbers introduced in Eq.(\ref{6}), it is quite evident in
Eqs.(\ref{x3}), (\ref{5}), (\ref{5A}), and as will be seen in more
detail below, that the basic numbers arise naturally,
automatically, in the theory of particles obeying interpolating
statistics. This feature is quite new and not recognized in the
standard theory of anyons in the literature. This might be an
indication that the theory of interpolating statistics naturally
involves a deformation of the system such as that described by the
basic number system with its consequences, without however
introducing a deformed algebra of operators.

\vspace{.15in}

 In addition to the representation,
\begin{equation}\label{6A}
    [n]= f^{n-1} + f^{n-3} + \cdots + f^{-n + 3 } + f^{-n +1} \, ,
\end{equation}
the following representation of the bracket number
\begin{equation}\label{7}
    [n] =  \frac{\sin n \pi \alpha }{\sin \pi \alpha}\, ,
\end{equation}
will be found most useful. In terms of the basic numbers, the
probabilities may now be expressed conveniently and succinctly as
\begin{eqnarray}
    P^{(2)}&=& \half (2 + [2]) \, , \nonumber \\
    P^{(3)} &=& \frac{1}{6}\left ( 6 + 7 [2] + 4[3] + [4]  \right
    ) \, , \nonumber \\
    P^{(4)} &=& \frac{1}{4 !} \left ( \, 35 + 54 [2] + 52 [3] +
    36 [4] + 18 [5]
    + 6 [6] + [7]  \,  \right )\label{8}\, ,
\end{eqnarray}
and so on.  We can derive a useful result,
\begin{equation}\label{9}
    [1] + [3] + [5] + \cdots + [2n-1]= \left (\, [n]\, \right  )^2\, .
\end{equation}
which can be proved by using \cite{GR} the identity,
\begin{equation}\label{10}
    \frac{1}{\sin t}\sum_{k=1}^n \, \sin(2k - 1)t\,
    = \left ( \frac{\sin nt}{\sin t} \right )^2\, .
\end{equation}
 We shall also find the following  result quite
useful:
\begin{equation}\label{11}
    [n-1]\, [n] = [2] + [4] + [6] + \cdots\, [2(n-1)]\, .
\end{equation}
In other words,  $ [2]\, [3] = [2] + [4], \, [3] \, [4] = [2] +
[4] + [6] , \,$ etc. This is proved by using the identity
\begin{equation}\label{12}
    \sum_{k=0}^{n-1}\; \sin k y = \sin \frac{(n-1) y}{2}\; \sin
    \frac{n y}{2}\; \csc \frac{y}{2}\, .
\end{equation}
These identities can be used to re-express Eqs.(\ref{8}) in the
form
\begin{eqnarray}
    P^{(2)}&=& \frac{1}{2}\left (          2 + [2] \right )\, ,
    \nonumber\\
P^{(3)}&=& \frac{1}{3 !}\left (2 +  [2]\right ) \left( 2 + 2 [2] +
[3] \right ) \, , \nonumber \\
    P^{(4)}&=& \frac {1}{4 !}\left ( 2+[2]\right )
    \left( 2+2[2]+[3]\right )
    \left ( 2+2[2] + 2[3] +[4] \right )\label{13}\, .
\end{eqnarray}
 In this manner,  generalizing to $n$ particles, we are led to
   the probability
  for the
 n-anyon state:
 \begin{equation}\label{14}
    P^{(n)}=\frac{1}{n \, !}
    \left ( 2+[2] \right ) \left ( 2 + 2[2] + [3]\right ) \cdots
    \left ( 2 + 2[2] + 2[3] + \cdots + 2 [n-1] + [n] \right )\, .
\end{equation}

\section{Detailed Balance}

From Eq.(\ref{14}),  we can infer the enhancement factor, which is
a measure of  how much greater  the  probability of the
$n+1\,$-particle state is, compared to the probability of the
$n$-particle state. For now, we simply express it as
\begin{equation}\label{15}
    F(n) =\frac{  P^{(n+1)}  } {P^{(n)}}\, ,
\end{equation}
which can be determined from Eq.(\ref{14}). This enhancement
factor provides the essential step in the method of Detailed
Balance. This brings us to the second assumption upon which our
theory of interpolating statistics rests: if $n_1, n_2$ represent
the average occupation numbers of states 1 and 2 respectively,
then the number of transitions flowing from 1 to 2 must equal that
flowing from 2 to 1 at equilibrium. This is the principle of
Detailed Balance. We should stress that this principle is
characteristic of thermodynamic equilibrium and may be regarded as
a consequence of the second law of thermodynamics \cite{thermo}.
Indeed the principle of Detailed Balance is valid when
thermodynamic equilibrium prevails or the validity of microscopic
reversibility in the language of statistical physics. This notion
is based on the reversibility of the microscopic equations of
motion, or on the Hermitian nature of the scattering Hamiltonian
\cite{Mattis}. The principle of Detailed Balance can thus be
stated as
\begin{equation}\label{16}
    n_1 F(n_2) e^{\ds \beta E_1} =  n_2 F(n_1) e^{\ds \beta E_2}\,
    ,
\end{equation}
where the population of each level is governed by the Boltzmann
factor and $F(n)$ is the enhancement factor. This yields
\begin{equation}\label{17}
    \frac{n}{F(n)} \; e^{\ds \beta E}= {\, \rm constant \;\;}= z\, ,
\end{equation}
which would, in principle, enable us to determine the distribution
function for the anyons in terms of the Boltzmann factor. In the
BE limit for instance, the enhancement factor is just $n+1$, which
immediately leads to the BE distribution
\begin{equation}\label{18}
    n= \frac{1}{   z^{-1} e^{\beta E}- 1    }\, .
\end{equation}
where $z = e^{\beta \mu}$ is the fugacity of the gas. Let us now
proceed in the same manner for arbitrary $f$. The enhancement
factor for arbitrary $f$ is given by Eqs.(\ref{14}) and
 (\ref{15}) which reduces to the simple form
\begin{equation}\label{19}
   F(n) = \frac{P^{(n+1)}}{P^{(n)}} =  \frac{1}{n+1}
    (2 + 2[2] +2[3]+ \cdots 2[n] + [n+1])\, .
\end{equation}

\vspace{.15in}

To proceed further, we employ the result
\begin{equation}\label{20}
    \sum_{k=0}^n \, [k]= 2 \cos (\pi \alpha /2)\, [n/2]\,
    [(n+1)/2]\, ,
\end{equation}
which is easy to prove by using identities involving sums of
trigonometric functions. This gives us the following result for
$F(n)$, after some algebra:
\begin{equation}\label{21}
    F(n)= \frac{1}{n+1} \,
    2 \cos (\pi \alpha /2) \, [(n+1)/2]\,
     \left \{\, [n/2]+ [n/2 +1]\,
    \right   \}\, .
\end{equation}
We can also derive another identity
\begin{equation}\label{22}
    [n/2]+ [n/2 +1]= 2 \cos (\pi \alpha /2) [(n+1)/2]\, ,
\end{equation}
which can be used to simplify the above result. Hence we obtain
\begin{equation}\label{23}
    F(n) = \frac{4}{n+1}\, \left \{[(n+1)/2] \, \cos (\pi \alpha /2)
    \right \}^2 \, .
\end{equation}
We observe that this reproduces the expected result $F(n)
\rightarrow n+1$ in the Bose limit. Upon now invoking the Detailed
Balance we obtain the important result
\begin{equation}\label{24}
    \frac{1}{z}e^{\beta E}= F(n)/n= \frac{4}{n(n+1)}
    \{[(n+1)/2]\, \cos (\pi \alpha /2)   \}^2\, .
\end{equation}
This can be rewritten in the following convenient form in order to
deal with the distribution function:
\begin{equation}\label{50C}
e^{\beta (E - \mu )}= \frac{1}{n (n+1)}\; \frac{\sin^2 (n+1)\pi
\alpha /2
    }
    {\sin^2 \pi \alpha /2} \, .
\end{equation}
Solving this equation should, in principle, lead to the
distribution function for the anyons in this formulation. At this
point before dealing with the distribution function, we need to
study the nature of the intermediate statistics as an
interpolation between the BE and FD limits. Specifically we need
to study the limits corresponding to BE and FD statistics.

\vspace{.15in}

 We have already observed that $F(n) \rightarrow n+1 $
 in the Bose limit,
 $f \rightarrow 1$. Indeed, it is readily verified that
Eq.(\ref{24}) reproduces the correct BE statistical distribution.
The case of Fermi limit, $\alpha \rightarrow 1, \, f \rightarrow
-1$, is, however, somewhat complicated. We have: $[n] \rightarrow
n, -n $ for odd and even occupation numbers respectively.
Evaluating the limit in Eq.(\ref{24}),   the enhancement factor
 may be put in the form,
\begin{equation}\label{25}
    \lim_{\alpha \rightarrow 1} F(n)= \frac{1}{n+1}\; \left ( {\cal I}m \; e^{i(n+1)\pi/2}
    \right )^2 \, ,
\end{equation}
thus resulting in a generalized Fermion theory. This corresponds
to an infinite dimensional representation analogous to the
generalized fermions investigated by Chaichian \emph{et al }.  If
we consider only $n=0,1, $ and impose the exclusion principle ``by
hand", then $F(n) \rightarrow 1, 0$. This is equivalent to
$F(n)\rightarrow 1-n $ and in this case it reproduces the standard
Fermi distribution with the exclusion principle. For arbitrary
$n$, which is allowed, however, it is an infinite value
representation
\begin{equation}\label{25A}
\lim_{\alpha \rightarrow 1}    F(n) = \frac{1}{n+1} \{1, 0, 1, 0,
1, \cdots \}\, ,
\end{equation}
and we find that $F(n) $ has repeating values $1/(n+1)$ and $ 0$
in the limit $ \alpha \rightarrow 1, f \rightarrow -1$. This is
thus a special and interesting feature of this theory. Thus for
arbitrary values of the parameter $f$, including the Fermi limit
$f \rightarrow -1$, the theory of intermediate statistics requires
the existence of generalized fermions beyond the exclusion
principle.

\vspace{.15in}

\section{The Distribution Function}

We begin by rewriting Eq.(\ref{50C}) in the form
\begin{equation}\label{30}
    e^{\beta (E - \mu )}= \frac{1}{n (n+1)}\; \frac{\sin^2 (n+1)
    x}
    {\sin^2 x} \, .
\end{equation}
Here $x = \pi \alpha /2$ in terms of the statistics determining
parameter. The object is to determine the average occupation
number $n$, the distribution function,  so that we can compare
with the standard BE or FD distribution and understand the nature
of the interpolating statistics. We can expand the right hand side
in a power series
\begin{equation}\label{31}
e^{\beta (E - \mu )} = \frac{1}{n} + a_0 + a_1 n + a_2 n^2 + a_3
n^3 + \cdots \, ,
\end{equation}
where
\begin{eqnarray}
a_0 &=& \csc^2 x (x \sin 2 x - \sin^2 x)\nonumber \\
a_1 &=& \csc^2 x (x^2 \cos 2 x - x \sin 2 x + \sin^2 x )
\nonumber \\
a_2 &=& \csc^2 x \left \{ - x^2 \cos 2x - \sin ^2 x + (x -
2x^3/3)\,
\sin 2x \right \}  \nonumber \\
 a_3 &=& \csc^2 x \left \{  (x^2-x^4/3)  \cos 2x + \sin
^2 x -(x - 2x^3/3)\, \sin 2x \right \} \label{32} \, ,
\end{eqnarray}
and so on. The coefficients to any desired order can be obtained
by using \emph{Mathematica}. It is clear that $a_0 \rightarrow 1,
\, $ while $a_n \rightarrow 0$ for $n\geq 1$ in the Bose limit,
which is consistent with the BE distribution as described in Sec.
III. In the case of the Fermi limit, $a_0 \rightarrow -1$ and it
leads to the generalized fermions. We can rewrite the above as
\begin{equation}\label{32A}
    \frac{1}{g}= n - \frac{a_1}{g}n^2 - \frac{a_2}{g}n^3 -
    \frac{a_3}{g}n^4 - \cdots \, .
\end{equation}
where $g = e^{\beta (E - \mu )}- a_0$. This series can be reverted
to express $n$ as a series in powers of $1/g $, thus
\begin{equation}\label{100A}
    n = 1/g + B/g^2 + C/g^3 + D/g^4 + E/g^5 + F/g^6 \cdots \, ,
\end{equation}
where
\begin{eqnarray}
  B &=& a_1/g,\; \; C=(2 a_1^2/g^2 + a_2/g ),\;
  D= (5 a_1 a_2/g^2 + a_3/g + 5 a_1^3/g^3) \, ,  \nonumber \\
  E&=& E= 6 a_1 a_3/g^2 + 3 a_2^2/g^2 +
  14 a_1^4/g^4 + a_4/g + 21 a_1^2
   a_2/g^3 \, , \nonumber \\
  F &=& 7 a_1 a_4/g^2 + 7 a_2 a_3/g^2 +
  84 a_1^3 a_2/g^4 + a_5/g + 28
  a_1^2 a_3/g^3 + 28 a_1 a_2^2/g^3 +
  42 a_1^5/g^5 \, ,    \label{101A}
\end{eqnarray}
and so on. Upon rearrangement,  we can rewrite the above form for
the distribution function as the following series:
\begin{eqnarray}
  n(E)&=& 1/g + a_1/g^3 + a_2/g^4 +
  (2 a_1^2 + a_3)/g^5  \nonumber \\
   &+& (5 a_1 a_2 + a_4)/g^6 +
   (5 a_1^3 + 6 a_1 a_3 + 3 a_2^2 +
   a_5)/g^7 + \cdots \, . \label{102A}
\end{eqnarray}

\vspace{.15in}

We observe that this is a power series in $1/g$ but the term with
$1/g^2$ is absent.  We can rewrite Eq.(\ref{102A}) in the form
\begin{equation}\label{40A}
    n(E)= 1/g + \sum_{k=3}^{\infty} \alpha_k/g^k\, ,
\end{equation}
where
\begin{equation}\label{41A}
    \alpha_3=a_1, \; \;   \alpha_4=a_2, \; \;  2 a_1^2 + a_3= \alpha_5,
   \;\;   5 a_1 a_2 + a_4 =  \alpha_6 \, ,
\end{equation}
etc. We can now express this in the form of a continued fraction.
We invoke the  standard algorithm which can be introduced as
follows \cite{Andrews}. If a continued fraction is of the form
\begin{equation}\label{50B}
\frac{B_n}{C_n}= c_0 + {b_1\over\displaystyle c_1 +
    {\strut b_2 \over\displaystyle c_2 +
     \cdots}}\, ,
\end{equation}
then the successive convergents (approximants) are obtained by the
recurrence formulae
\begin{equation}\label{51B}
    B_n = c_n B_{n-1} + b_n B_{n-2}, \quad C_n = c_n C_{n-1}
    + b_n C_{n-2},\quad B_{-1}= 1, \; C_{-1}= 0\, .
\end{equation}
Thus in order to put the series in Eq.(\ref{40A}) in the continued
fraction form,  we proceed with the first convergent and set $c_0
= B_0/C_0 = 0$. Using the recurrence relations, Eq.(\ref{51B}), we
determine $B_0 = 0, C_0 = 1, b_1 = 1, c_1 = g, B_1=1, C_1=g$. This
determines the first convergent as $n = 1/g$. Next, the recurrence
relations lead to $b_2 = -\alpha_3 g, c_2 = g^2 + \alpha_3, B_2 =
g^2 + \alpha_3, C_2 = g^3$ which determines the second convergent
to be
\begin{equation}\label{52B}
n=  {1\over\displaystyle g -
    {\strut \alpha_3 g  \over\displaystyle g^2 + \alpha_3
     }}\, .
\end{equation}
Continuing on with successive convergents in this manner, we
obtain the desired infinite continued fraction form for the
distribution function as follows
\begin{equation}\label{42B}
    n(E)= {1\over\displaystyle g -
    {\strut \alpha_3 g \over\displaystyle g^2 + \alpha_3 -
    {\strut \alpha_4 g \over \displaystyle \alpha_3 g + \alpha_4 -
    {\strut \alpha_5 g \over \displaystyle \alpha_4 g + \alpha_5
    - \cdots}}}}
\end{equation}
This is the distribution function in the exact theory, albeit not
in a closed form. Approximations can be implemented in order to
investigate the thermostatistical properties of the anyons. The
first approximant,
\begin{equation}\label{43A}
    n(E)= \frac{1}{e^{\beta (E - \mu)} - a(\alpha)}\, ,
\end{equation}
was indeed employed, as an  ansatz, in our earlier investigation
\cite{RAPN}. It might be pointed out that the continued fraction
form is mathematically desirable in view of the fact that it can
be expressed as a ratio of polynomials, hence analytic in the
variable $\alpha$ as well as offering the best approximation
possible. This form is also amenable to investigation in terms of
Pade approximants.

\vspace{.15in}

 Finally it must be pointed out that Eq.(\ref{42B})
is an exact expression for the distribution function, albeit in
the form of an infinite continued fraction, and can be determined
explicitly for any specific value of the parameter $\alpha$,
either as an infinite continued fraction, or to any desired
approximation. For instance, for the case of $\alpha = \half$, the
exact form of the distribution function is
\begin{equation}\label{50B}
    n(E)|_{\alpha = \half}= {1\over\displaystyle g +
    {\strut a_0 g \over\displaystyle g^2 - a_0 +
    {\strut (a_0 - \pi^3/48 ) g \over\displaystyle a_0 g -(a_0 -
    \pi^3/48) +
     \cdots}}}\, ,
\end{equation}
where $a_0(\half )= \pi/2 - 1$.

\section{Summary}

The thermodynamic distribution function of anyons in two space
dimensions, obeying interpolating statistics  is not known in the
literature and is an open question. We have not only found an
answer, thus determining the distribution function, we have also
demonstrated that the theory describing interpolating statistics
has several remarkable and interesting features. Our investigation
leads to a generalized definition of permutation symmetry in
arbitrary dimensions and not restricted to two space dimensions.
We have shown that the theory of permutation symmetry that would
describe particles obeying interpolating statistics is succinctly
formulated in the language of  basic numbers. These basic numbers
arise naturally and automatically in this formulation but do not
explicitly invoke any deformed oscillator algebra. Our theory is
based on the principle of Detailed Balancing which  is  a
consequence of the Second law of thermodynamics. Furthermore, this
formulation leads to the determination of the exact distribution
function, without having to introduce any approximation. In this
manner, we have formulated the theory which leads to a
transcendental equation for the distribution function of anyons in
terms of the statistics determining parameter and the Boltzman
factor containing the energy of the state. We obtain a solution of
this equation which we express as a power series as well as in the
form of a continued fraction. We show that the first approximation
of this theory reproduces a form of the distribution function
introduced by us in an earlier investigation.

\vspace{.15in}

 An important feature of our formulation
consists of the fact that the basic numbers arise naturally and
automatically in this theory, specifically the symmetric
formulation of the basic numbers. The basic numbers arise
automatically in our theory but they are not part of an oscillator
algebra, nor do we introduce the construction of Fock space for
the particles obeying the intermediate statistics. The theory
reduces to BE statistics in the Bose limit, $f \rightarrow 1,
\alpha \rightarrow 0$. The Fermi limit, $f \rightarrow -1, \alpha
\rightarrow 1$ leads to generalized fermions beyond the exclusion
principle and thus correspond to infinite dimensional
representation and they reduce to the familiar FD statistics only
when $n$ is restricted to 0, 1 by hand.

\vspace{.15in}

Although the algebra of q-oscillators appears in the literature on
the subject of anyons \cite{Frau,Frappat,Chai}, it is also known
that q-oscillators may have nothing to do with anyons since the
former exist in arbitrary space-time dimensions. In our
formulation, we do not use q-oscillator algebra, neither do we use
Fock states. We do not make use of the Hamiltonian other than to
recognize that the Hamiltonian should be of a form that permits
Detailed Balance. The Hamiltonian consists of only the kinetic
energy terms and is the free Hamiltonian \cite{Lerda} in the anyon
gauge and the anyon wave functions satisfy ``twisted" boundary
conditions.

\vspace{.15in}

It is interesting that the first approximant (first convergent) of
our solution corresponds to the approximate form for the
distribution function
\begin{equation}\label{104A}
    n(E)= \frac{1}{e^{\beta E - \mu)} - a(\alpha)}\, ,
\end{equation}
which is the form introduced in an earlier investigation
\cite{RAPN}. It is interesting that this also agrees with the
partition function derived by the method of Green function
\cite{Chai}, to first order approximation. It should be stressed
that while the form above is a point of agreement for an
approximate theory, the form of $a$ in the present formulation as
in Eq.(\ref{32}), as a function of the statistics determining
parameter, is very different from that of \cite{RAPN}. Finally we
obtain the exact form of the distribution function for the case
$\alpha = \half$ in the form of a continued fraction which can be
evaluated to any desired order. Such a form for the distribution
function can indeed be obtained for any chosen value of $\alpha$,
the statistics determining parameter

\vspace{.15in}

It is of much interest that this formulation leads to an
intermediate statistical mechanics, interpolating between BE and
FD, without explicit use of the special properties valid in 2+1
space-time dimensions, but only stems from the principle of
Detailed Balance. This contrasts with investigations in the
literature of anyons which arise from the braid group in the
special case of 2+1 space-time. As is seen from Eq.(\ref{30}), the
dependence on the statistics determining parameter is through a
quadratic function, hence invariant under $\alpha \rightarrow -
\alpha$, implying a clockwise-counterclockwise symmetry of the
braid \cite{Khare}. It would be worthwhile to continue this
investigation further in order to explore the relation to these
anyons. As this formulation reveals several desirable and
interesting theoretical features, it is appropriate to pose some
questions such as the following. Does the generalization to any
space dimensions imply that the connection with Chern-Simons type
of gauge theory can be extended beyond two space dimensions? Is it
possible to formulate the partition problem corresponding to this
generalized, exact theory? Is it possible to determine the many
useful thermodynamic properties of the system described by such an
exact theory? These questions will occupy us in further
investigations of this theory.

\end{document}